\documentclass[useAMS,usenatbib]{mn2e}
\usepackage{graphicx}
\usepackage[section]{placeins}
 
\usepackage{times}

\begin{document}

\title[Asteroids in retrograde resonance]{Asteroids in retrograde resonance with Jupiter and Saturn}
\author[M. H. M. Morais and F. Namouni]{M. H. M. Morais$^{1}$\thanks{E-mail:
helena.morais@ua.pt (MHMM); namouni@obs-nice.fr (FN)} and  F.
Namouni$^{2}$\footnotemark[1]\\
$^{1}$Department of Physics \& I3N, University of Aveiro, Campus Universit\'ario de Santiago, 3810-193 Aveiro, Portugal \\
$^{2}$Universit\'e de Nice, CNRS, Observatoire de la C\^ote d'Azur, BP 4229, 06304 Nice, France}

\date{}

\maketitle

\begin{abstract}
We identify a set of  asteroids among Centaurs and Damocloids, that orbit contrary to the common direction of motion in the Solar System and that enter into resonance with Jupiter and Saturn. Their orbits have inclinations $I\ga140^\circ$ and semi-major axes $a<15$~AU. Two objects are  currently in retrograde resonance with Jupiter: 2006 BZ8 in the 2/-5 resonance  and 2008 SO218 in the 1/-2 resonance. One object, 2009 QY6, is currently in the 2/-3 retrograde resonance with Saturn.  These are the first  examples of Solar System objects in retrograde resonance.  The present resonant configurations last for several thousand years. Brief captures in retrograde resonance with  Saturn are also possible during the $20{,}000$ years integration timespan, particularly in the 1/-1 resonance (2006 BZ8) and the 9/-7 resonance (1999 LE31).
\end{abstract}

\begin{keywords}
celestial mechanics; minor planets, asteroids;  comets: general; Oort Cloud; Kuiper belt
\end{keywords}

\section{Introduction}
The discovery of exoplanets that revolve around their host stars in the opposite direction to stellar rotation has renewed interest in the dynamics of retrograde motion in $N$-body gravitational systems. In the Solar System,  all major and most minor planets have prograde orbits around the Sun except for 50 small bodies known to orbit opposite the common direction (Minor Planet Center as of 1 July 2013). Such bodies are thought to originate mainly in the Oort Cloud comet reservoir \citep{Jewitt2005AJ}. However, numerical simulations have recently indicated that retrograde orbits in the inner solar system may be produced through the gravitational excitation exerted by the gas giants  \citep{Greenstreet_etal2012ApJ}. Understanding the dynamics of retrograde minor bodies may provide clues to the mechanisms that reverse major and minor planets' orbits alike. In this Letter, we seek to determine the dynamical states of the minor bodies that reside  on retrograde orbits in the vicinity of Jupiter and Saturn. We show that some of the known minor bodies may be trapped temporarily in mean motion resonances with these planets where the ratio of their orbital periods are integer fractions. 

Unlike a number of extrasolar multi-planet systems ($\sim 20$\%), the Solar System's planets are not in resonance even though  the pairs Jupiter-Saturn and Neptune-Uranus are close to the 5/2 and 2/1 mean motion resonances respectively. For minor bodies, the role of mean motion resonances is complex. For instance,  the  3/1 mean motion resonance with Jupiter in the main asteroid belt is a known source of Near Earth Asteroids \citep{Wisdom1985,Morbidelli_etal2002}. On the other hand, the Hilda asteroid family resides stably at the 3/2 resonance with Jupiter \citep{Nesvorny&FerrazMello1997}. Mean motion resonances also protect objects on Neptune-crossing orbits in the Kuiper belt from disruptive close encounters with this planet \citep{Morbidelli_etal1995,Malhotra1996}. The existence of small bodies on retrograde orbits in resonance with the giant planets may provide another example of the dynamical diversity and the roles played by resonant excitation. 

In section 2, we recall the main features of retrograde mean motion resonances. In section 3, we present our numerical simulations and identify the various observed dynamical states. Section 4 contains a summary and discussion of the significance of these findings.

\section{Retrograde resonance}
In order to test whether a particular asteroid in retrograde motion around the Sun is in resonance with a planet, we need to identify the correct resonant terms of the standard Fourier expansion of the three-body problem's disturbing function  \citep{SSDbook}. Keeping with the usual definition of the osculating orbital elements where retrograde orbits are inclined by $I>90^\circ$ and $\dot\lambda>0,\ \dot\lambda^\prime>0$ where $\lambda$ and $\lambda^\prime$ are the asteroid's and the planet's mean longitudes respectively, we show elsewhere \citep{Morais&Namouni2013} that the resonant terms for the p/-q resonance are of the form:
\begin{equation}
\phi= q \lambda- p \lambda^\prime-(p+q-2k)\varpi+2\,k\,\Omega,  \label{resangle}
\end{equation}
where $p$, $q$ and $k$ are integers, $p+q\ge2\,k$ and the longitude of pericentre is  $\varpi=\omega-\Omega$, where   $\omega$ and $\Omega$ are respectively the argument of pericentre and  the longitude of ascending node\footnote{When motion is retrograde, the usual ascending node becomes the descending node. To conserve the canonical nature of the longitude of pericenter, that is measured from the true ascending node, its expression becomes $\varpi=\omega-\Omega$.  The d'Alembert rule is not obeyed in (\ref{resangle}) because for the retrograde orbit, angles are measured in a direction contrary to the motion of the prograde orbit.}.  To lowest order in eccentricity and inclination, the corresponding force amplitude is proportional to $\cos^{2k}(I/2) e^{p+q-2\,k}$. When the inclination is close to $180^\circ$ (nearly coplanar retrograde motion) the dominant resonant term has $k=0$ and amplitude proportional to $e^{p+q}$.  This reflects the fact that a p/-q retrograde resonances is weaker than its p/q prograde counterpart (whose force is proportional to $e^{|p-q|}$) as an  encounter of an asteroid and a planet orbiting in opposite directions around the Sun occurs at a higher relative velocity during a shorter time than in a prograde configuration. We note that the definition of the p/-q resonant angle may be expressed using orbital angle variables that assume a retrograde longitude (e.g. $\dot\lambda>0,\ \dot\lambda^\prime<0$) along with an inclination $I<90^\circ$  \citep{Morais&Giuppone2012}.  We prefer the use of (\ref{resangle}) as we have developed a simple approach to identify analytical resonant amplitudes from the classic expression of the disturbing function for prograde orbits (see \citet{Morais&Namouni2013}). Ultimately, resonance is independent of the coordinates' choice as long as they are obtained through a suitable canonical transformation from the usual orbital elements.

\section{Numerical integrations of objects in retrograde orbits}
To identify retrograde mean motion resonances with Jupiter and Saturn, we searched the Minor Planet Center  database\footnote{http://www.minorplanetcenter.net/} for objects in retrograde orbits with semi major axes $a<15$~AU and inclinations $I\ga140^\circ$ that were observed at multiple oppositions (uncertainty parameter $\le 2$).  The inclination constraint  restricts the sample to nearly co-planar retrograde orbits thereby simplifying the identification of mean motion resonances. Fig.~1 shows the distributions of $(a,I)$ and $(a,e)$ for the selected objects. Table 1 shows  their orbital  elements at JD2456400.5 obtained from JPL/HORIZONS\footnote{http://ssd.jpl.nasa.gov}.

The first object in Table I is  Near Earth Object 343158 (2009 HC82) an  Apollo family member that always remains interior to Jupiter's orbit. The other five objects cross Jupiter's orbits and  are sometimes referred to as Centaurs\footnote{A Centaur is a  small body on a long-term unstable giant-planet crossing orbit with $5.5<a<30.1$~AU as defined by JPL$^3$. The Minor Planet Center$^2$ has a stricter definition such that the pericentre of a Centaur's orbit must be exterior to Jupiter's orbit.}. 
 Whereas  2005 VD has a pericentre just inside Jupiter's orbit but does not cross Saturn's orbit,  1999 LE31, 2008 SO218 and 2006 BZ8 cross both Jupiter's and Saturn's orbits, and 2009 QY6  also crosses Uranus' orbit.  These objects have high eccentricity and/or high inclination orbits, resembling Halley-type comet orbits. With a Tisserand parameter
$T<2$, these objects may be part of the Damocloid population \citep{Jewitt2005AJ} whose physical properties indicate that their likely source is the Oort Cloud similarly to Halley-type comets. The source of Centaurs with  low to moderate inclinations is likely to be the Kuiper belt \citep{Volk&Malhotra2013}.

\begin{figure}
  \centering
    \includegraphics[width=8.6cm]{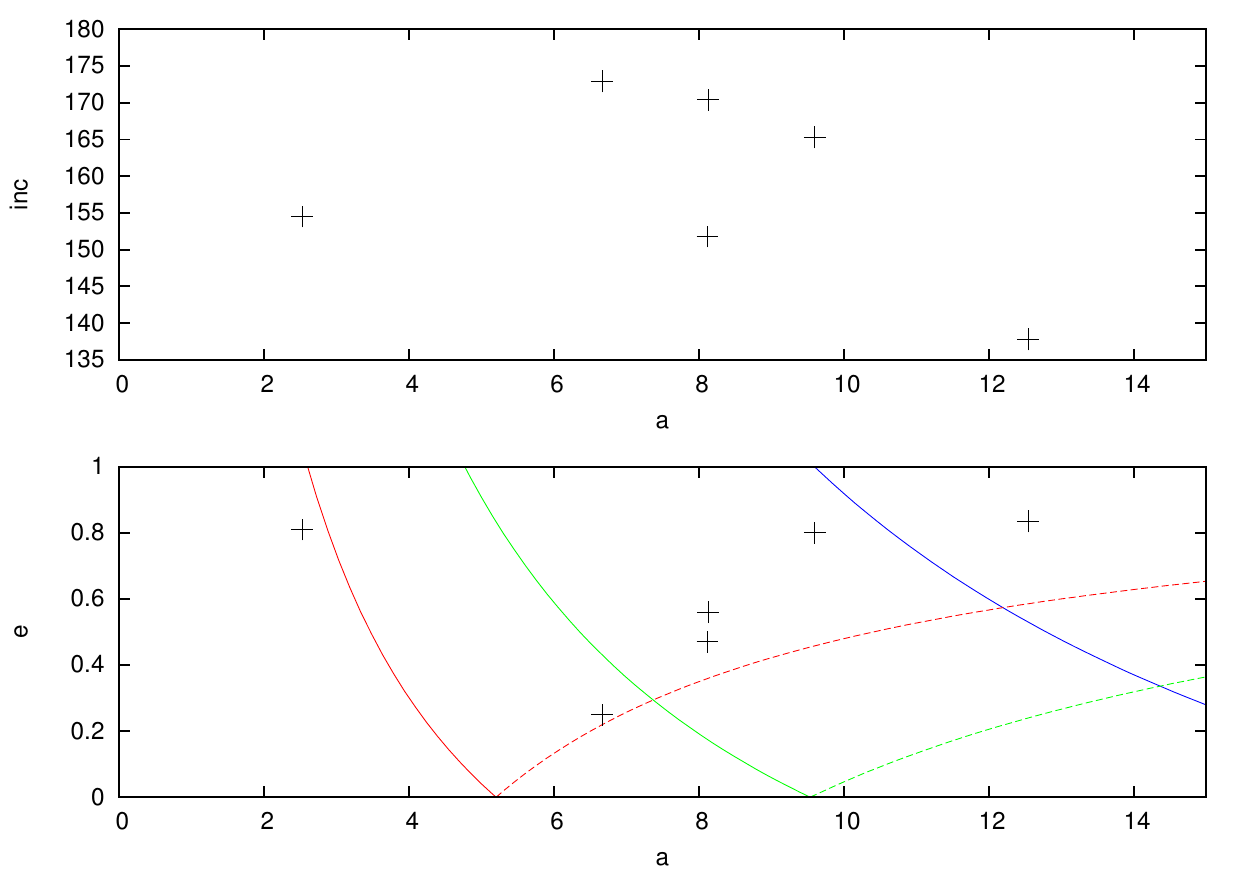}
\caption{Distributions of $(a,I)$ and $(a,e)$ for our subset of retrograde asteroids.}    
\end{figure}

We numerically integrated  massless objects with orbital elements from Table I using  the Burlisch-Stoer integrator of the package MERCURY \citep{Chambers1999} with an accuracy parameter $10^{-12}$.  The main eight planets were included and initial conditions (nominal orbits) for all bodies were obtained from JPL/HORIZONS at  JD2456400.5.   We  generated 9 clones for each object with orbital elements sampled from a 6-dimensional gaussian distribution around the nominal orbit. The standard deviations (1-$\sigma$ uncertainties) of the orbital elements were obtained from ASTDYS\footnote{http://hamilton.dm.unipi.it/astdys/}.   We followed the nominal and clones' orbits for $\pm10{,}000$~yrs centered on JD2456400.5.

\begin{table*}
\begin{center}
\begin{tabular}{|c||c|c|c|c|c|c|c|}
\hline
                  & $a$ (AU) &  $e$  &  $I$ ($^\circ$)  &  $M$ ($^\circ$) & $\omega$ ($^\circ$) & $\Omega$ ($^\circ$) & $T$ \\
\hline
343158 (2009 HC82)   & 2.5277835(3) & 0.8074783(2) & 154.49716(2) & 38.93235(5) & 298.51511(7) & 294.93344(8) & 1.317\\
2005 VD       & 6.67019(8)& 0.250215(8) & 172.8732(1) & 147.325(4) & 179.080(3) & 173.2478(3) & -1.395  \\
1999 LE31    & 8.1226(3) & 0.46824(2) & 151.80672(4) & 223.18(1) &   32.323(2)  & 291.9894(2) & -1.314 \\
330759 (2008 SO218) & 8.13497(2) & 0.564005(1)  & 170.35962(1) &  51.1362(2) & 354.59079(9) & 348.10764(9) & -1.396 \\
2006 BZ8      & 9.60667(6) & 0.800888(1) & 165.29992(1) &  82.2981(8)  & 82.31622(5)	&  183.6132(8) & -1.032  \\
2009 QY6      & 12.541(1) & 0.83473(1) & 137.75983(6) &  28.649(4)  & 195.0817(2) &  197.62243(4)  & -0.851 \\
\hline
\end{tabular}
\caption{Nominal orbital elements and Tisserand parameter of the selected retrograde asteroids at JD2456400.5, taken from HORIZONS with ASTDYS 1-$\sigma$ uncertainties (in brackets) on the last significant decimal place.}
\end{center}
\end{table*}

\begin{figure}
  \centering
    \includegraphics[width=8.6cm]{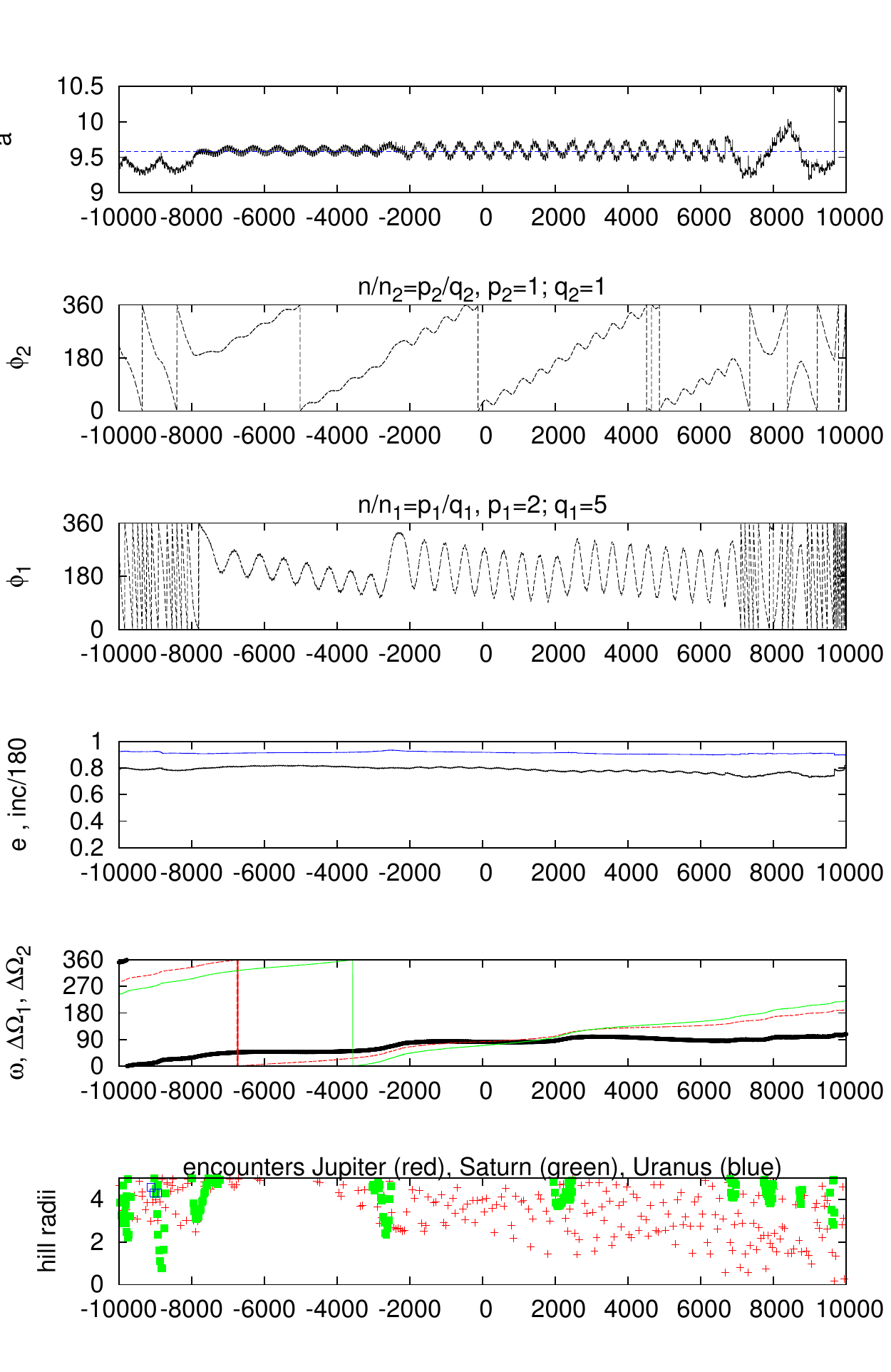}
\caption{Nominal orbit of 2006 BZ8 (from top to lower panel): semi-major axis $a$ and location of $p_1/q_1$ resonance with inner planet (Jupiter); $\phi_2$ is angle of $p_2/q_2$ resonance with outer planet (Saturn); $\phi_1$ is angle of $p_1/q_1$ resonance with inner planet (Jupiter); eccentricity $e$ (lower curve in black) and inclination $inc$ divided by $180^\circ$ (upper curve in blue);  argument of pericenter $\omega$ (black thick curve) and  difference in nodal longitudes between asteroid and Jupiter (red curve) or Saturn (green curve); encounters with Jupiter (red plus), Saturn (green filled square) and Uranus (blue open square) in units of Hill's radii of these planets.}   
\label{2006BZ8} 
\end{figure}
\begin{figure}
  \centering
    \includegraphics[width=8.6cm]{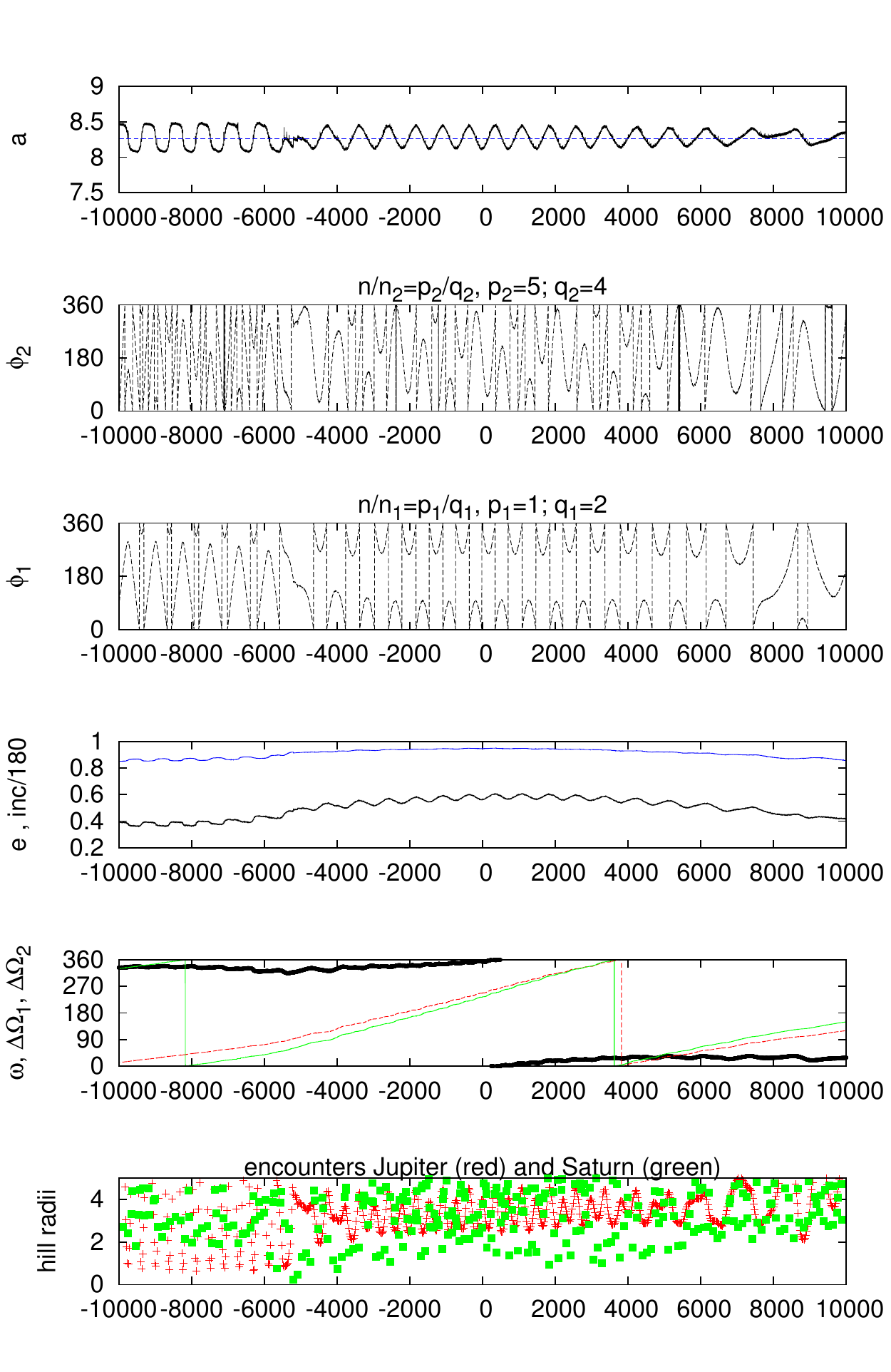}
\caption{Cloned orbit of 2008 SO218. Plotted quantities are those of Fig.~2. The departure from the nominal orbit is a combined 1-$\sigma$ deviation is each orbital element. The cloned orbit was chosen to present an example of asymmetric libration.}
\label{2008SO218}    
\end{figure}
\begin{figure}
  \centering
    \includegraphics[width=8.6cm]{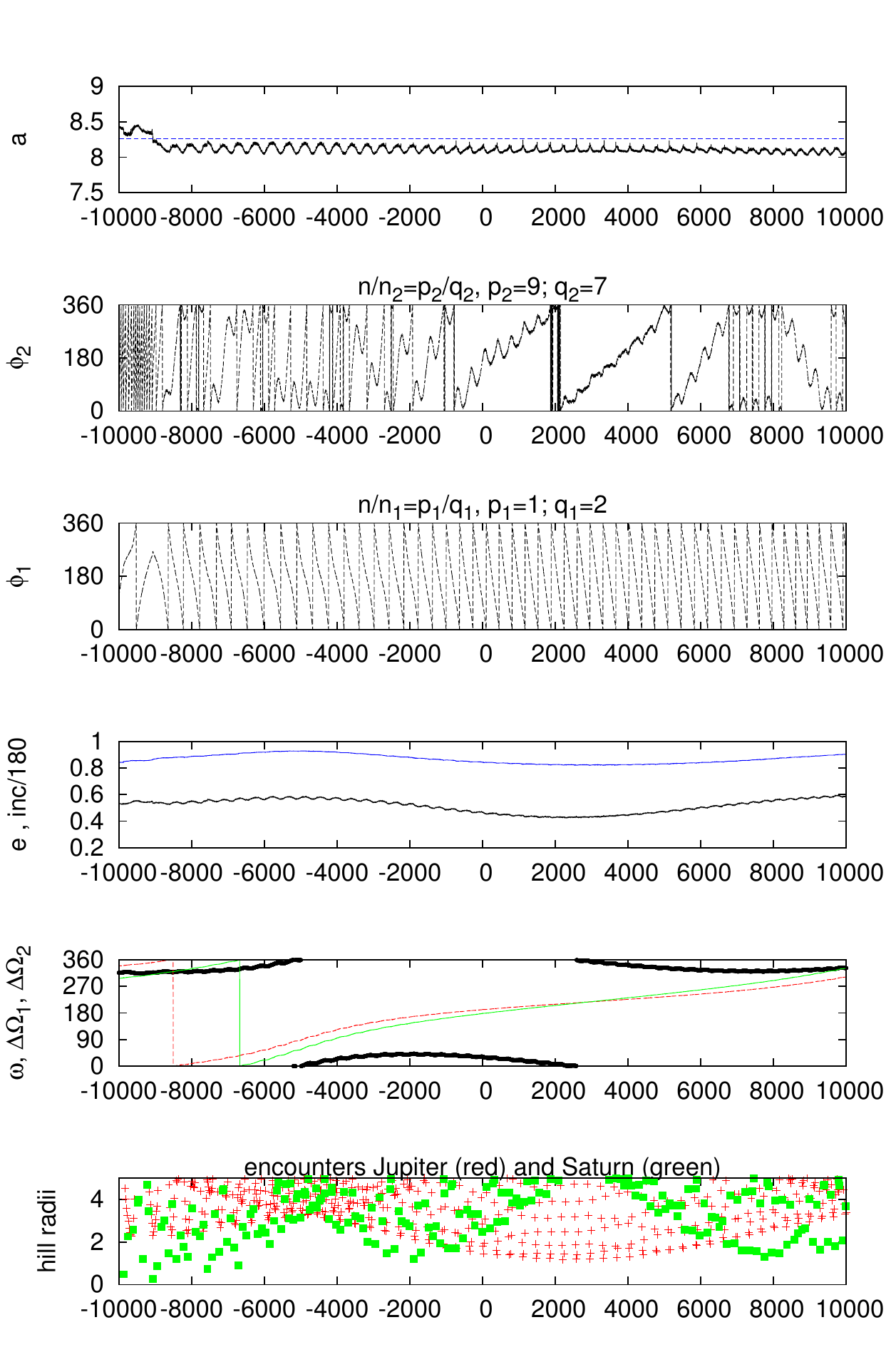}
\caption{Nominal orbit of 1999 LE31. Plotted quantities are those of Fig.~2. }    
\label{1999LE31}
\end{figure}
\begin{figure}
  \centering
    \includegraphics[width=8.6cm]{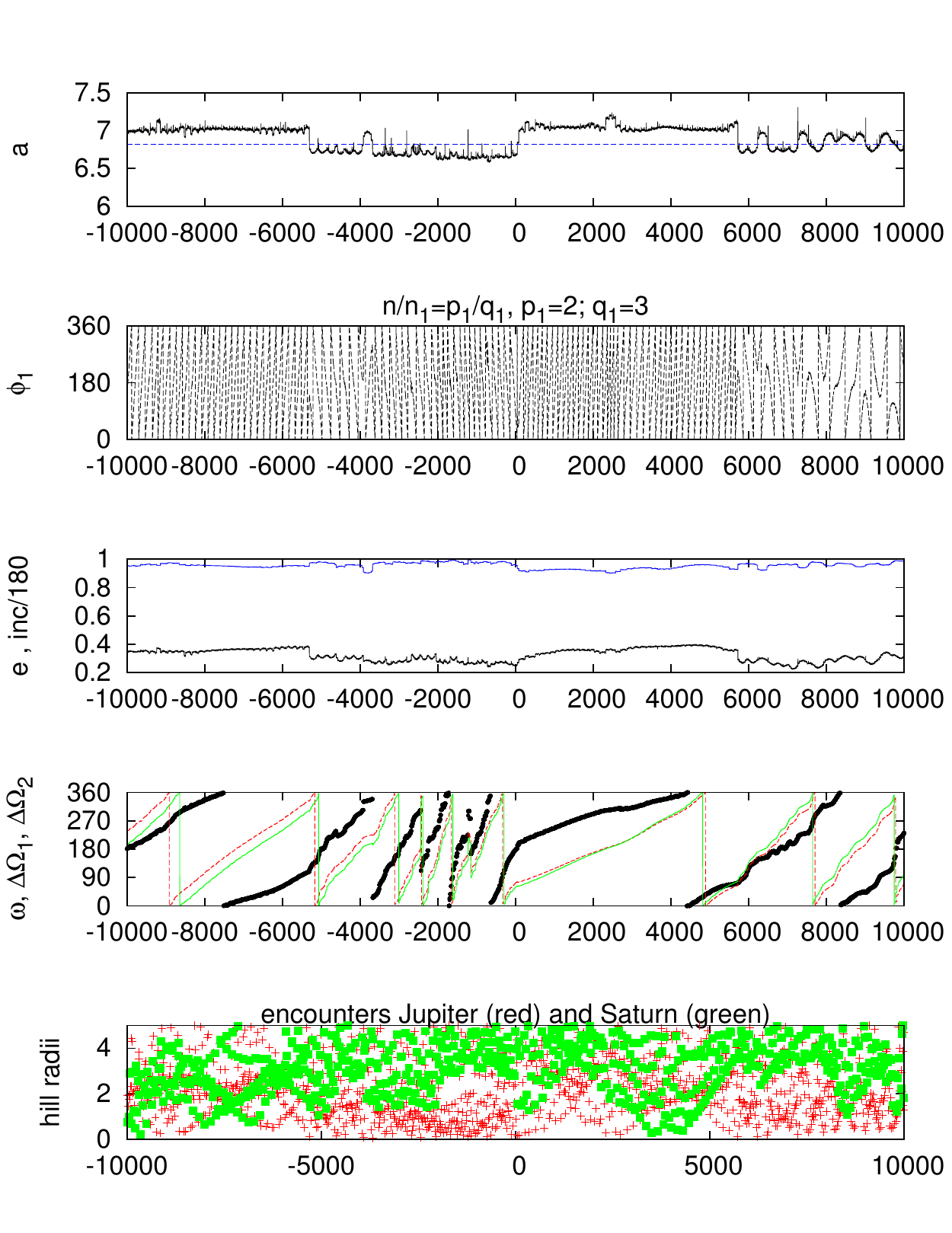}
\caption{Nominal orbit of 2005 VD. Plotted quantities are those of Fig.~2.}   
\label{2005VD} 
\end{figure}
\begin{figure}
  \centering
    \includegraphics[width=8.6cm]{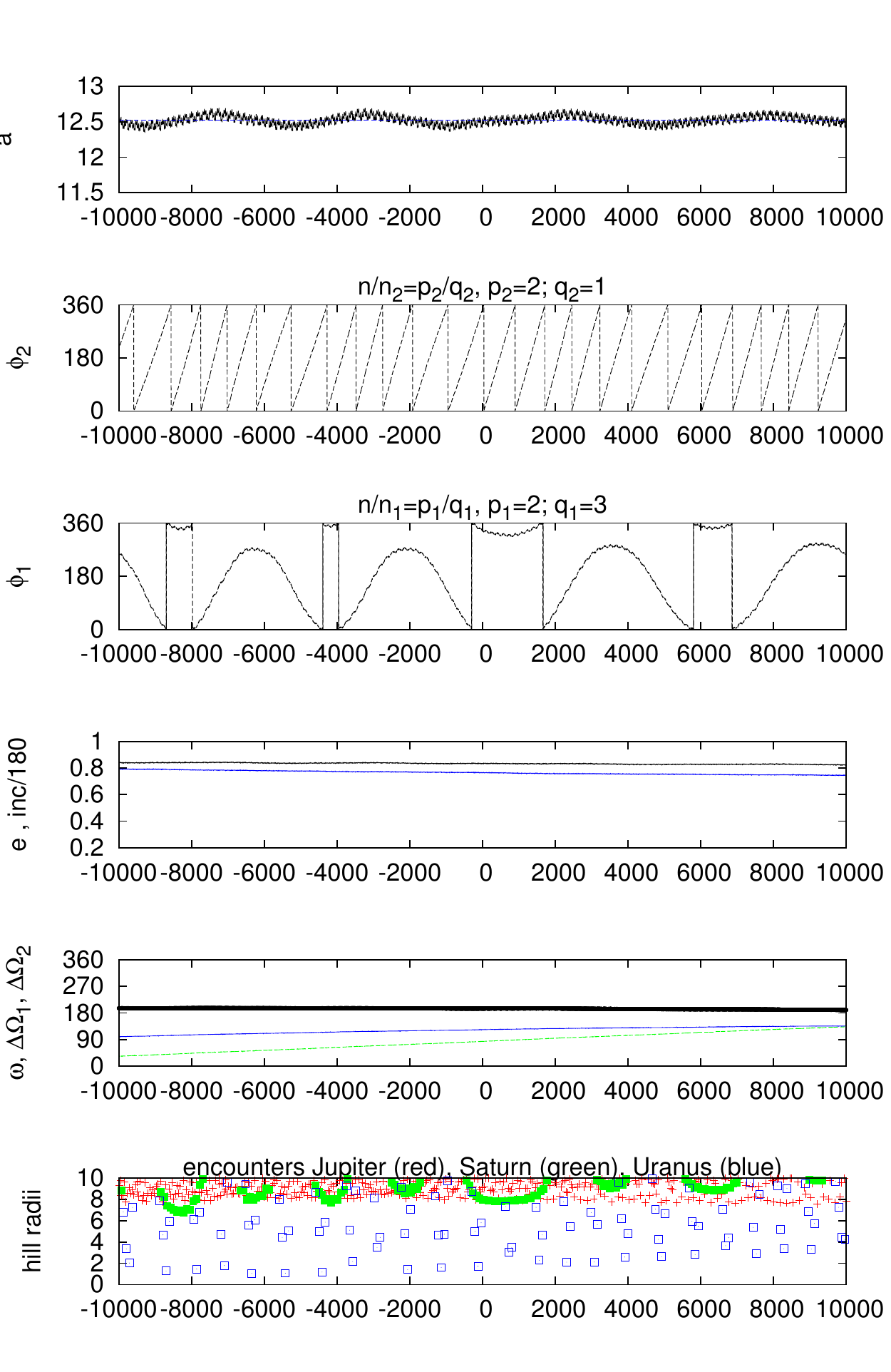}
\caption{Nominal orbit of 2009 QY6. Plotted quantities are those of Fig.~2 but inner planet is Saturn and outer planet is Uranus. Also shown the difference in nodal longitudes with Saturn (green) or Uranus (blue) }   
\label{2009QY6} 
\end{figure}
\begin{figure}
  \centering
    \includegraphics[width=8.6cm]{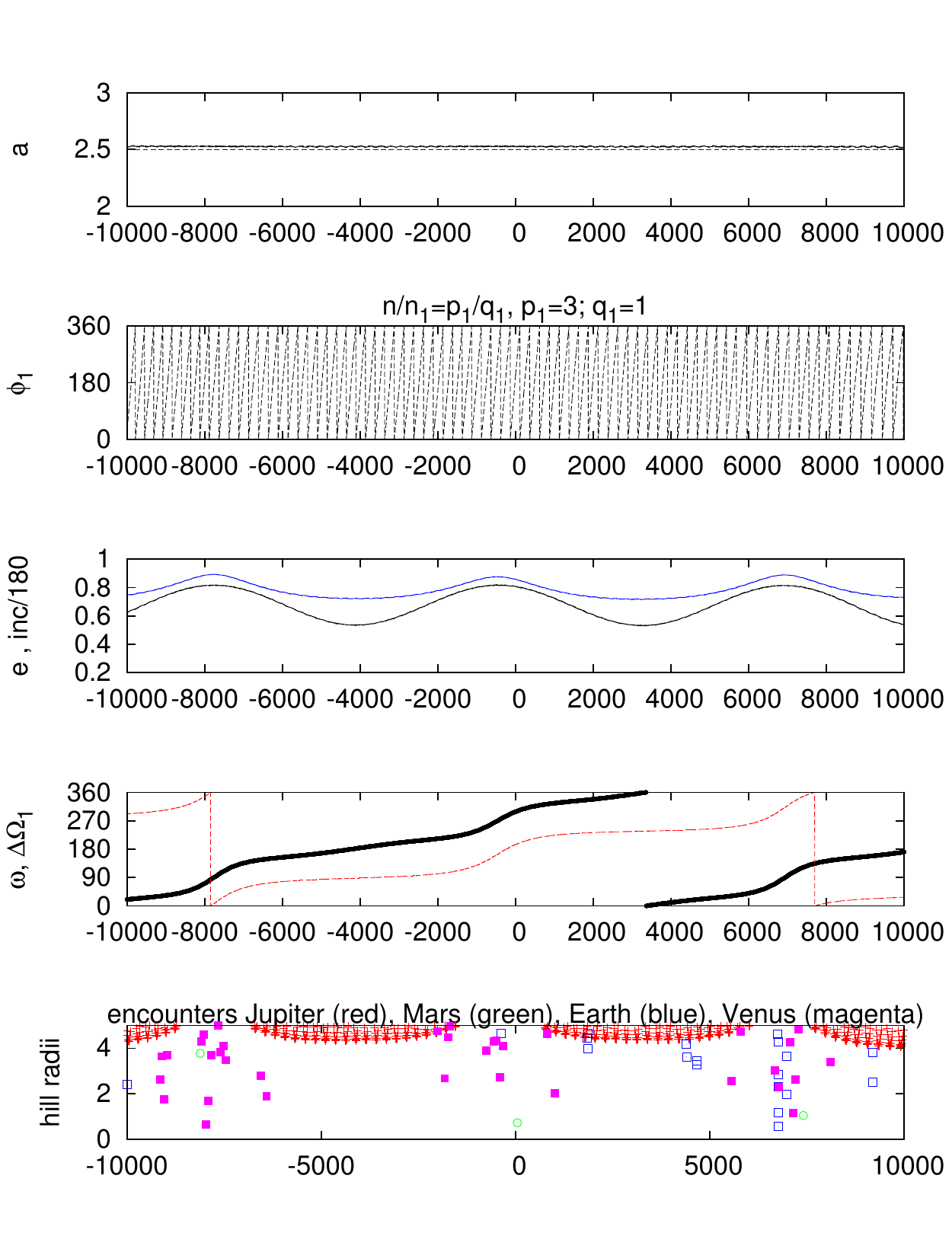}
\caption{Nominal orbit of 2009 HC82. Plotted quantities are those of Fig.~2  but outer planet is Jupiter and appropriate values of the resonance integers $p_1$ and $q_1$ are used. Also shown encounters with Jupiter (red plus), Mars (green open circle), Earth (blue open square) and Venus (magenta filled square) in units of hill's radii of these planets} 
\label{2009HC82}   
\end{figure}

2006 BZ8  (Fig.~\ref{2006BZ8}) is currently in  the 2/-5  resonance  with Jupiter (libration around $180^\circ$) while it is near the  1/-1 resonance with Saturn (circulation).  All clones show similar behavior with  capture in the present configuration occurring at  $-8{,}000$~yrs and an exit at around $7{,}000$~yrs. While the 2/-5 resonance with Jupiter lasts, the distance to Saturn remains larger than 2 Hill's radii.  The 2/-5 resonance with Jupiter protects the asteroids from close approaches to this planet for a while, despite the orbit's  large eccentricity. The distance to Jupiter remains larger than 1 Hill's radius until around $7{,}000$~yrs when very close encounters with this planet occur. As a consequence, the asteroid exits the 2/-5 resonance with Jupiter but a  brief capture in the 1/-1 resonance with Saturn is possible (1/10 clones) as shown  between $7{,}000$~yrs and $10{,}000$~yrs in this example\footnote{We note that we count the nominal orbit as one of the clones.}. The argument of pericentre $\omega$ remains close to 0, indicating a Kozai secular resonance\footnote{We confirmed libration of $\omega$ with longer term integrations.}.

2008 SO218 (Fig.~\ref{2008SO218}) is currently in the 1/-2 resonance with Jupiter (libration around 0) while it is near the  5/-4 resonance with Saturn.  When the 1/-2 resonant angle librates around 0 the asteroid is protected from close approaches to Jupiter  and the distance to this planet remains larger than 2  Hill's radii, but it can experience close approaches with Saturn.  Although all clones exhibit currently symmetric libration around 0, episodes of asymmetric libration in the 1/-2 resonance are also possible (6/10 clones) during which the asteroid can have close approaches to Jupiter, as shown between $-10{,}000$~yrs and $-5{,}000$~yrs in this example.  All clones stay near the 1/-2 resonance with Jupiter during  the integration timespan. The argument of pericentre $\omega$ remains near 0, indicating a Kozai secular resonance$^7$.

1999 LE31 (Fig.~\ref{1999LE31}) is currently near the 1/-2 resonance with Jupiter while it is near the  9/-7 resonance with Saturn. All clones show similar behavior with present circulation of the 9/-7 resonant angle with Saturn  but brief libration is possible (9/10 clones) as shown between $7{,}000$~yrs and $9{,}000$~yrs in this example. Close approaches to Saturn at around $-8{,}000$ yrs brought the asteroid to the vicinity of the 9/-7 resonance with this planet. The distance to Jupiter remains larger than 1 Hill's radius. The argument of pericentre $\omega$ executes a Kozai libration around $0$.$^7$ 
 
2005 VD (Fig.~\ref{2005VD}) has its current pericentre very close to Jupiter's orbit, while its apocentre is close to Saturn's orbit. It is also the retrograde asteroid closest to the solar system's median plane with an largest inclination ($I=172.9^\circ$).\footnote{Only minor planet 2013 LA2's uncertain orbit exceeds such inclination with  $I=175^\circ$ (MPC uncertainty parameter $=6$).} It therefore suffers repeated close encounters with Jupiter and Saturn, making its orbit  very chaotic as can be seen through the  fast semi-major axis drift. Accordingly, the clones' dynamical behavior varies widely and any resonance capture is short lived. In this example, the asteroid  passes the 7/-11, 9/-13, 7/-11 and finally the 2/-3 resonances with Jupiter. 

2009 QY6 (Fig.~\ref{2009QY6}) is currently in the 2/-3 resonance with Saturn (large amplitude asymmetric libration) while it is near the 2/-1 resonance with Uranus.  The 2/-3 resonance configuration can last all integration timespan (2/10 clones) or at least half the integration timespan (7/10 clones). Passage through the 2/-3 resonance with Saturn near the current time (without capture) is also possible (1/10 clones). The distance to Jupiter and Saturn remains larger than about 7 Hill's radii. The orbit is moderately chaotic due to close encounters with Uranus. The argument of pericentre $\omega$ executes a Kozai libration around $180^\circ$.$^7$

2009 HC82 (Fig.~\ref{2009HC82}) is currently located near the 3/-1  resonance with Jupiter (the resonant argument circulates) and has very regular behavior despite experiencing close encounters with Venus, Earth and Mars. 
 The parameter $h=(1-e^2)\cos^2{I}=0.28$ hence the orbit's evolution is probably dominated  by the  Kozai secular resonance with  Jupiter \citep{Kinoshita&Nakai2007} that forces the slow circulation of the argument of pericentre $\omega$. The origin of this asteroid is likely to be related to its proximity to the 3/1 resonance with Jupiter as numerical integrations by \citet{Greenstreet_etal2012ApJ} indicate that  this  resonance can a be source of retrograde near-Earth asteroids.

\section{Conclusion}
This analysis is to our knowledge the first identification of Centaurs and Damocloids currently in retrograde resonance with the giant planets. Although long-lived Centaurs are known to hop between prograde resonances with the planets \citep{Bailey&Malhotra2009},  there was no previous report of retrograde resonant capture. The relevant resonances observed in this Letter are the 2/-5 resonance with Jupiter for 2006 BZ8 (libration around $180^\circ$), 1/-2 resonance with Jupiter for 2008 SO218 (libration around 0), and the 2/-3 resonance with Saturn for 2009 QY6  (asymmetric libration). The present resonant configurations last for several thousand years. Moreover, retrograde resonant capture with Saturn is possible within the $20{,}000$ year integration timespan for  2006 BZ8 (1/-1), and  for 1999 LE31 (9/-7).  The identified resonant configurations  are temporary as the minor planets experience close encounters with the giant planets. This is akin to the temporary capture of NEAs in the (prograde) 1/1 resonance \citep{Namouni_etal1999PhRvL,Christou2000} with the terrestrial planets and  that of Centaurs in the same resonance with Uranus and Neptune\citep{FuenteMarcos2012A&A,FuenteMarcos2013A&A,Alexandersen_etal2013}. 

As our studied asteroids may be termed as Damocloids (Tisserand parameter $T<2$), they too are likely to originate in the Oort cloud as nuclei of dead comets. This conclusion was reached by \citet{Jewitt2005AJ} based on the physical study of 20 such objects. However with the exception of 1999 LE31 \citep{Jewitt2005AJ} and 2005 VD \citep{Pinilla_etal2013A&A}, the physical properties of our asteroid sample have not been studied so far. A possible lack of ultrared matter, rich organic material that cannot survive in the inner solar system, must await photometric analysis of the asteroids in resonance to confirm their similarity or lack thereof with the known Damocloid population \citep{Sheppard2010AJ}. If such objects are indeed the nuclei of dead comets that entered the giant planet region on initially retrograde orbits then their semi-major axis drift rate in that region must be particularly small in order to be trapped in retrograde resonances. The reason is that retrograde resonances are much weaker than their prograde counterparts as physical encounters occur at larger velocities during a shorter time \citep{Morais&Namouni2013}. The strongest retrograde resonance is the co-orbital 1/-1 resonance (of order 2 in eccentricity) yet it did not dominate resonant capture in our sample. This indicates that the dynamics of retrograde resonant capture may be more complex and could involve an interplay of retrograde mean motion resonances with the secular Kozai resonance  \citep{Gomes_etal2005,Gallardo_etal2012} as  observed for 2006 BZ8, 2008 SO218, 1999 LE31 and 2009 QY6.  Another interesting possibility is given by the recent numerical integrations of \citet{Greenstreet_etal2012ApJ} who observed an orbital inversion in the inner solar system. That work indicates that  the 3/1 prograde mean motion resonances may produce large inclination near Earth asteroids like 2009 HC82. It is not yet clear whether the prograde resonance may provide a preferential path to retrograde resonance capture. Further  dynamical and physical study of the minor planets identified in this Letter will  help discriminate between the possible origins of retrograde resonance capture.

\section*{Acknowledgments}
We thank the anonymous referee for helpful suggestions that improved the article.
We acknowledge financial support from FCT-Portugal (PEst-C/CTM/LA0025/2011).

\bibliographystyle{mn2e}

\bibliography{retrograde}

\begin{thebibliography}{}

\bibitem[\protect\citeauthoryear{{Alexandersen}, {Gladman}, {Greenstreet},
  {Kavelaars} \& {Petit}}{{Alexandersen} et~al.}{2013}]{Alexandersen_etal2013}
{Alexandersen} M.,  {Gladman} B.,  {Greenstreet} S.,  {Kavelaars} J.~J.,
  {Petit} J.-M.,  2013, ArXiv e-prints

\bibitem[\protect\citeauthoryear{{Bailey} \& {Malhotra}}{{Bailey} \&
  {Malhotra}}{2009}]{Bailey&Malhotra2009}
{Bailey} B.~L.,  {Malhotra} R.,  2009, \icarus, 203, 155

\bibitem[\protect\citeauthoryear{{Chambers}}{{Chambers}}{1999}]{Chambers1999}
{Chambers} J.~E.,  1999, \mnras, 304, 793

\bibitem[\protect\citeauthoryear{{Christou}}{{Christou}}{2000}]{Christou2000}
{Christou} A.~A.,  2000, \icarus, 144, 1

\bibitem[\protect\citeauthoryear{{de la Fuente Marcos} \& {de la Fuente
  Marcos}}{{de la Fuente Marcos} \& {de la Fuente
  Marcos}}{2012}]{FuenteMarcos2012A&A}
{de la Fuente Marcos} C.,  {de la Fuente Marcos} R.,  2012, \aap, 547, L2

\bibitem[\protect\citeauthoryear{{de la Fuente Marcos} \& {de la Fuente
  Marcos}}{{de la Fuente Marcos} \& {de la Fuente
  Marcos}}{2013}]{FuenteMarcos2013A&A}
{de la Fuente Marcos} C.,  {de la Fuente Marcos} R.,  2013, \aap, 551, A114

\bibitem[\protect\citeauthoryear{{Gallardo}, {Hugo} \& {Pais}}{{Gallardo}
  et~al.}{2012}]{Gallardo_etal2012}
{Gallardo} T.,  {Hugo} G.,    {Pais} P.,  2012, \icarus, 220, 392

\bibitem[\protect\citeauthoryear{{Gomes}, {Gallardo}, {Fern{\'a}ndez} \&
  {Brunini}}{{Gomes} et~al.}{2005}]{Gomes_etal2005}
{Gomes} R.~S.,  {Gallardo} T.,  {Fern{\'a}ndez} J.~A.,    {Brunini} A.,  2005,
  Celestial Mechanics and Dynamical Astronomy, 91, 109

\bibitem[\protect\citeauthoryear{{Greenstreet}, {Gladman}, {Ngo}, {Granvik} \&
  {Larson}}{{Greenstreet} et~al.}{2012}]{Greenstreet_etal2012ApJ}
{Greenstreet} S.,  {Gladman} B.,  {Ngo} H.,  {Granvik} M.,    {Larson} S.,
  2012, \apjl, 749, L39

\bibitem[\protect\citeauthoryear{{Jewitt}}{{Jewitt}}{2005}]{Jewitt2005AJ}
{Jewitt} D.,  2005, \aj, 129, 530

\bibitem[\protect\citeauthoryear{{Kinoshita} \& {Nakai}}{{Kinoshita} \&
  {Nakai}}{2007}]{Kinoshita&Nakai2007}
{Kinoshita} H.,  {Nakai} H.,  2007, Celestial Mechanics and Dynamical
  Astronomy, 98, 67

\bibitem[\protect\citeauthoryear{{Malhotra}}{{Malhotra}}{1996}]{Malhotra1996}
{Malhotra} R.,  1996, \aj, 111, 504

\bibitem[\protect\citeauthoryear{{Morais} \& {Giuppone}}{{Morais} \&
  {Giuppone}}{2012}]{Morais&Giuppone2012}
{Morais} M.~H.~M.,  {Giuppone} C.~A.,  2012, \mnras, 424, 52

\bibitem[\protect\citeauthoryear{{Morais} \& {Namouni}}{{Morais} \&
  {Namouni}}{2013}]{Morais&Namouni2013}
{Morais} M.~H.~M.,  {Namouni} F.,  2013, Celest. Mech Dyn. Astr., submitted,
  arXiv preprint arXiv:1305.0016

\bibitem[\protect\citeauthoryear{{Morbidelli}, {Bottke} Jr., {Froeschl{\'e}} \&
  {Michel}}{{Morbidelli} et~al.}{2002}]{Morbidelli_etal2002}
{Morbidelli} A.,  {Bottke} Jr. W.~F.,  {Froeschl{\'e}} C.,    {Michel} P.,
  2002, Asteroids III, pp 409--422

\bibitem[\protect\citeauthoryear{{Morbidelli}, {Thomas} \&
  {Moons}}{{Morbidelli} et~al.}{1995}]{Morbidelli_etal1995}
{Morbidelli} A.,  {Thomas} F.,    {Moons} M.,  1995, \icarus, 118, 322

\bibitem[\protect\citeauthoryear{{Murray} \& {Dermott}}{{Murray} \&
  {Dermott}}{1999}]{SSDbook}
{Murray} C.~D.,  {Dermott} S.~F.,  1999, {Solar system dynamics}.
{Cambridge University Press}

\bibitem[\protect\citeauthoryear{{Namouni}, {Christou} \& {Murray}}{{Namouni}
  et~al.}{1999}]{Namouni_etal1999PhRvL}
{Namouni} F.,  {Christou} A.~A.,    {Murray} C.~D.,  1999, Physical Review
  Letters, 83, 2506

\bibitem[\protect\citeauthoryear{{Nesvorny} \& {Ferraz-Mello}}{{Nesvorny} \&
  {Ferraz-Mello}}{1997}]{Nesvorny&FerrazMello1997}
{Nesvorny} D.,  {Ferraz-Mello} S.,  1997, \icarus, 130, 247

\bibitem[\protect\citeauthoryear{{Pinilla-Alonso}, {Alvarez-Candal}, {Melita},
  {Lorenzi}, {Licandro}, {Carvano}, {Lazzaro}, {Carraro}, {Al{\'{\i}}-Lagoa},
  {Costa} \& {Hasselmann}}{{Pinilla-Alonso} et~al.}{2013}]{Pinilla_etal2013A&A}
{Pinilla-Alonso} N.,  {Alvarez-Candal} A.,  {Melita} M.~D.,  {Lorenzi} V.,
  {Licandro} J.,  {Carvano} J.,  {Lazzaro} D.,  {Carraro} G.,
  {Al{\'{\i}}-Lagoa} V.,  {Costa} E.,    {Hasselmann} P.~H.,  2013, \aap, 550,
  A13

\bibitem[\protect\citeauthoryear{{Sheppard}}{{Sheppard}}{2010}]{Sheppard2010AJ}
{Sheppard} S.~S.,  2010, \aj, 139, 1394

\bibitem[\protect\citeauthoryear{{Volk} \& {Malhotra}}{{Volk} \&
  {Malhotra}}{2013}]{Volk&Malhotra2013}
{Volk} K.,  {Malhotra} R.,  2013, \icarus, 224, 66

\bibitem[\protect\citeauthoryear{{Wisdom}}{{Wisdom}}{1985}]{Wisdom1985}
{Wisdom} J.,  1985, \nat, 315, 731

\end{thebibliography}

\end{document}